\documentclass[lettersize,journal]{IEEEtran}
\usepackage{amsmath,amssymb,amsfonts}
\usepackage{algorithmic}
\usepackage{graphicx}
\usepackage{algorithm,algorithmic}
\usepackage{hyperref}
\usepackage{textcomp}

\usepackage{booktabs, subcaption}   
\usepackage[style=ieee]{biblatex} 
\def\BibTeX{{\rm B\kern-.05em{\sc i\kern-.025em b}\kern-.08em
    T\kern-.1667em\lower.7ex\hbox{E}\kern-.125emX}}
\begin{document}
\title{Real-time Neonatal Chest Sound Separation using Deep Learning}
\author{Yang Yi Poh, \IEEEmembership{Student Member, IEEE}, Ethan Grooby, \IEEEmembership{Student Member, IEEE}, \\Kenneth Tan, Lindsay Zhou, Arrabella King, Ashwin Ramanathan, Atul Malhotra, \\Mehrtash Harandi, Faezeh Marzbanrad \IEEEmembership{Senior Member, IEEE}
\thanks{E. Grooby acknowledges the support of the Monash University Postgraduate Publication Award. A. Malhotra's research is supported by funding from the Cerebral Palsy Alliance, and the National Health and Medical Research Council, Australia.}
\thanks{Y. Poh, E. Grooby, M. Harandi and F. Marzbanrad are with the Department of Electrical and Computer Systems Engineering, Monash University, Melbourne, Australia. K. Tan, L. Zhou, A. King, A. Ramanathan and A. Malhotra are with Monash Newborn, Monash Children's Hospital and Department of Paediatrics, Monash University, Melbourne, Australia. E. Grooby is with the BC Children's Hospital Research Institute and the Department of Electrical and Computer Engineering, University of British
Columbia, Vancouver, Canada.}
\thanks{email: Yang.Poh@monash.edu}}

\maketitle

\begin{abstract}
Auscultation for neonates is a simple and non-invasive method of providing diagnosis for cardiovascular and respiratory disease. Such diagnosis often requires high-quality heart and lung sounds to be captured during auscultation. However, in most cases, obtaining such high-quality sounds is non-trivial due to the chest sounds containing a mixture of heart, lung, and noise sounds. As such, additional preprocessing is needed to separate the chest sounds into heart and lung sounds. This paper proposes a novel deep-learning approach to separate such chest sounds into heart and lung sounds. Inspired by the Conv-TasNet model, the proposed model has an encoder, decoder, and mask generator. The encoder consists of a 1D convolution model and the decoder consists of a transposed 1D convolution. The mask generator is constructed using stacked 1D convolutions and transformers. The proposed model outperforms previous methods in terms of objective distortion measures by 2.01\,dB to 5.06\,dB in the artificial dataset, as well as computation time, with at least a 17 times improvement. Therefore, our proposed model could be a suitable preprocessing step for any phonocardiogram-based health monitoring system.
\end{abstract}


\section{Introduction}
\label{sec:introduction}
\IEEEPARstart{A}{uscultation} for neonatal care is a critical component of physical examinations. It provides access to heart and lung sounds, which can be used to diagnose cardio-respiratory conditions and monitor vital signs. Its application ranges from regular heart-rate assessment \cite{springer_robust_2014}, \cite{liu_open_2016} to computer-aided diagnosis \cite{song_diagnostic_2021}, \cite{sung_computer-assisted_2015}. However, these algorithms work best when using high-quality heart or lung sounds, while heart and lung sounds typically only come in pairs and are contaminated by noise. As such, further processing is needed to isolate the individual sources of sound.

There are challenges when separating pure heart and lung sounds in newborns: (a) Newborns typically have weak heart and lung sounds due to their smaller organ size. (b) Typical newborn heart sounds have a frequency band between 50\,Hz and 250\,Hz, while newborn lung sounds have a frequency band between 200\,Hz and 1000\,Hz \cite{nersisson_heart_2017}, causing an overlap in their spectrum. (c) Newborns have a smaller chest area and as such, focusing the auscultation for the desired heart or lung sound is more difficult. (d) High levels of noise in the environment such as crying sounds and respiratory support sounds can interfere with the obtained chest sound mixture.

Traditionally, many chest sound separation methods require heart sound segmentation and lung sound segmentation. Table~\ref{tab:prev_method} summarises some of the methods used. For heart sound segmentation, these methods typically identify the first heart sound (S1) and the second heart sound (S2). However, most of these methods struggle in a high-noise scenario. Our recent works showed that using Non-Negative Factorisation (NMF) and Non-Negative Co-Factorisation (NMCF) outperforms these traditional methods when performing chest sound separation in newborn children \cite{grooby_noisy_2022}. Despite that, there are still some limitations. Namely, computation cost and the performance in the presence of respiratory support noise remain weaknesses of the method.

\begin{table*}[htbp]
  \caption{Previous methods used for performing single-channel sound separation.}
  \label{tab:prev_method}
  \begin{tabular}{p{0.2\linewidth}  p{0.7\linewidth}}
    \toprule
    Method & Description\\ 
    \midrule
    Adaptive Fourier Decomposition & Based on the energy distribution, adaptive Fourier decomposition is used to clean identified S1 and S2 heart sound peaks.\\
    Adaptive Line Enhancement & Applies adaptive filtering on the original and time-delayed recording to extract semi-periodic components (heart sounds).\\
    Empirical Mode Decomposition (EMD)& Decomposes signals into a sum of oscillatory functions called intrinsic mode functions (IMFs). For each IMF, S1 and S2 peaks are identified and filtered to obtain heart and lung components. \\
    Filtering & Passband frequencies of 50-250\,Hz and 200-1000\,Hz were used to obtain heart and lung sounds, respectively.\\
    Interpolation & Identified S1 and S2 heart sound peaks are removed either by 20-300\,Hz bandstop filter or complete elimination of the sections. Interpolation in the time-frequency domain is then performed to recover the lung sounds.\\
    Modulation Filtering & Involves bandpass and bandstop filtering of temporal trajectories of the short-term spectral components, to obtain heart and lung sounds respectively. \\
    NMF clustering 1 and 2 & Both methods blindly decompose the mixture into numerous sub-components. All components are then clustered into either heart or lung based on spectral or temporal criteria. \\
    Recursive Least Squares Adaptive Filtering & Identified S1 and S2 heart sound peaks are used to create a reference heart sound. The recursive least squares filter then uses the original recording and the reference heart sound to obtain clean heart and lung sounds. \\
    Singular Spectrum Analysis (SSA) & SSA decomposes the signal into principal components. The top eigenvector pairs of components with the strongest frequency component less than 250\,Hz are assigned as heart sounds.\\
    Wavelet Transform-based Filter & Wavelet threshold is used to separate stationary sounds (heart sounds) from non-stationary sounds (heart sounds) using adaptive threshold with adjusting multiplicative factor\\
    Wavelet Decomposition and SSA & Wavelet decomposition is performed to obtain a 0-500\,Hz signal. SSA is then performed to obtain just heart sounds.\\ 
    \bottomrule
  \end{tabular}
\end{table*}

Recently, deep learning-based audio source separation has been proposed in a variety of domains. With the success of deep neural networks, they are the state of the art when it comes to supervised separation. As a result, domains with large datasets such as those in the speech domain \cite{luo_conv-tasnet_2019, luo_dual-path_2020, huang_joint_2015} and music domain \cite{sawata_all_2021, kavalerov_universal_2019, defossez_hybrid_2022} are dominated by deep neural networks. However, when it comes to chest sounds, from neonatal to adult, a smaller amount of data is available. If the training data is too small, supervised separation would cause overfitting, thus reducing the overall performance of the model. As such, many different approaches have been proposed to overcome this limitation. For instance, Wang et at. used NMF to aid in the deep learning process \cite{wang_heart-lung_2023}, while Tsai et al. exploited the periodicity of heart and lung sounds to perform the separation \cite{tsai_blind_2020}. Adding to this, data augmentation-based learning will be explored in this paper to artificially increase the number of samples and reduce overfitting.

In addition, deep learning-based audio source separation models have been dominated by either convolutional neural networks (CNN) \cite{defossez_hybrid_2022, tsai_blind_2020} or long short-term memory (LSTM) networks \cite{luo_dual-path_2020, huang_joint_2015}. Typically, CNNs have the advantage of capturing local features well. However, CNNs must be sufficiently deep to capture a desired receptive field. LSTM networks, on the other hand, are capable of learning long-term dependencies. Nonetheless, LSTM networks can suffer from exploding and vanishing gradients, making it difficult to train. In recent times, state-of-the-art audio encoders have adopted a transformer architecture due to their excellent ability to model sequential data \cite{radford_robust_2022, chung_w2v-bert_2021}. As such, a transformer-based network architecture will be explored in this paper.

The rest of the paper will be organised as follows. We first introduce the model architecture in Section \ref{sec:model_architecture}, describe the data acquisition, experimental configuration, and performance evaluation criteria in Section \ref{sec:experiments}, and present the experimental results in Section \ref{sec:results}.

\subsection*{Contributions}
The main contributions of this paper are as follows:
\begin{enumerate}
    \item Present a convolution and transformer-based neural architecture to perform blind source separation.
    \item Present a data augmentation strategy to increase the variety of samples during training.
    \item Demonstrate that our model outperforms the latest neonatal chest sound separation methods, in objective distortion measures, heart rate and breathing rate estimation, and computation time. 
    \item Providing our full model publicly available, for future application integration and comparison: \href{https://github.com/yangyipoh/Neonatal-Chest-Sound-Separation-using-Deep-Learning}{https://github.com/yangyipoh/Neonatal-Chest-Sound-Separation-using-Deep-Learning}.
\end{enumerate}

\section{Model Architecture}
\label{sec:model_architecture}
\begin{figure}[ht]
  \centering
  \includegraphics[width=0.8\linewidth]{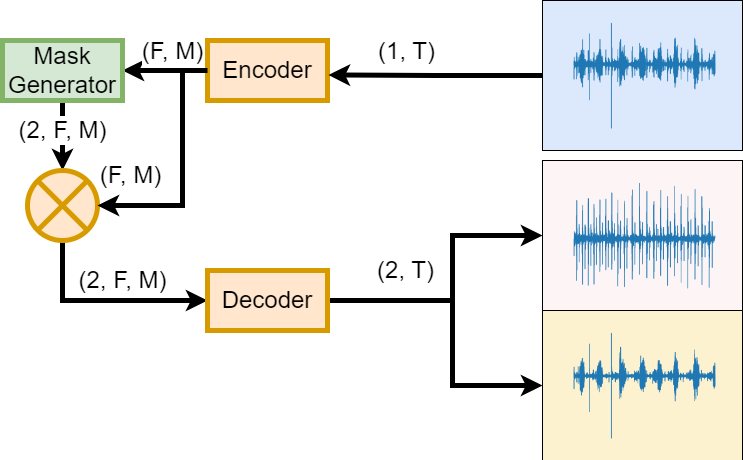}
  \caption{The model architecture of the proposed model}
  \label{fig:model}
\end{figure}

Inspired by the Conv-TasNet model\cite{luo_conv-tasnet_2019}, the model architecture is broken down into 3 components: encoder, decoder, and mask generator. Figure~\ref{fig:model} shows the overall system block diagram. The encoder first segments the mixture waveform by sliding a windowing function and encodes the segment as an n-dimensional feature space. This turns the input waveform of size $(1, T)$ into a 2-dimensional feature space of size $(F, M)$, where $F$ represents the feature dimension and $M$ represent the number of frames or number of hops. The mask generator then takes this 2-dimensional feature space and produces $s$ feature space masks, where $s$ is the number of desired sources in the mixture. The mask generated would have a shape of $(s, F, M)$. In this case, the number of desired sources is 2 for heart and lung sounds, and the mask generator would produce a mask with a shape of $(2, F, M)$. Each mask is then applied to the feature space and then passed to the decoder to transform from the feature space back to the waveform. 

\begin{figure}[ht]
    \centering
    \includegraphics[width=0.8\linewidth]{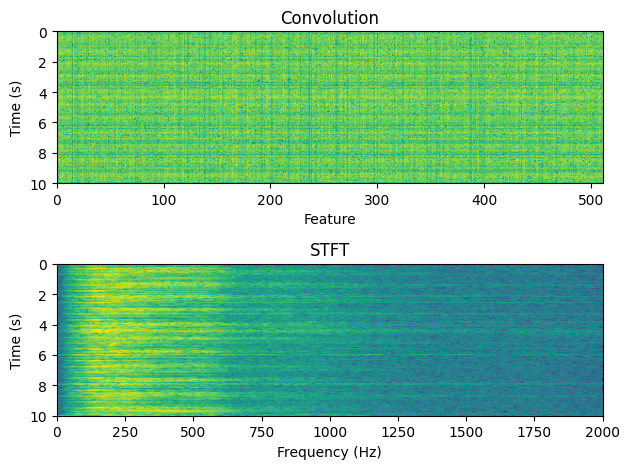}
    \caption{Comparison between using 1D convolution vs. Short-Time Fourier Transform (STFT) as an encoder.}
    \label{fig:stft_vs_conv}
\end{figure}

The encoder and decoder are represented by a 1D convolution and a 1D transposed convolution. Traditionally, time-frequency representations such as Short-Time Fourier Transform (STFT) and Mel Spectrogram are chosen, where the feature space contains information about the frequency content. However, there are some issues with this implementation: (1) There is no guarantee that time-frequency is the optimal representation for the task of chest sound separation, especially when there is frequency overlap between the sources. (2) Because the signal of interest only exists in a relatively narrow frequency band (50\,Hz to 250\,Hz for heart, 200\,Hz to 1000\,Hz for lung \cite{nersisson_heart_2017}), an STFT representation would be inefficient as little to no information is provided at the higher frequency contents. Instead, a learnable 2-dimensional feature space is preferred as the model is able to optimise the feature space based on the dataset. Figure~\ref{fig:stft_vs_conv} shows an example of a convolution representation compared to an STFT representation for a typical clean chest sound. Here, it can be seen that the convolution representation better utilizes the feature space compared to the STFT representation.

\begin{figure}[ht]
  \includegraphics[width=\linewidth]{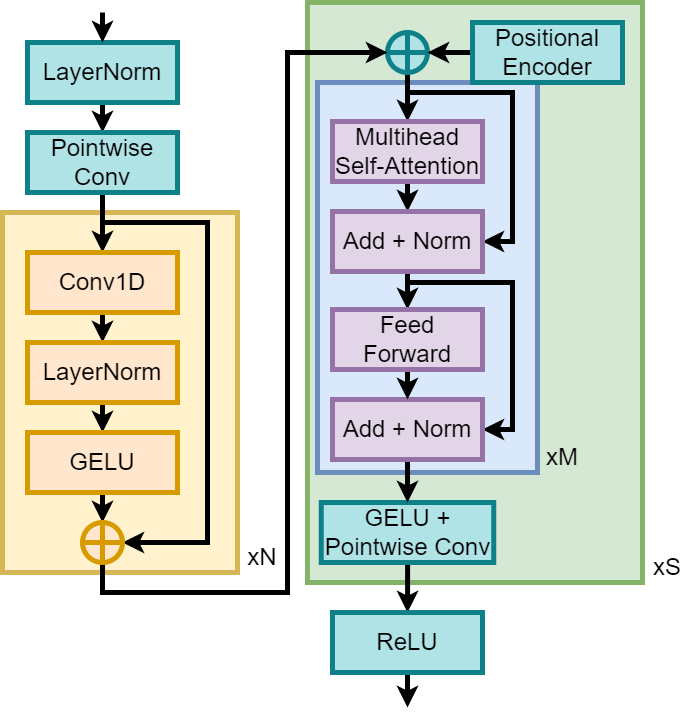}
  \caption{The mask generator in the model. The mask generator takes in the feature space of shape $(F, M)$ and produces $s$ feature space masks of shape $(s, F, M)$.}
  \label{fig:mask_generator}
\end{figure}

Where the proposed model differs from its inspiration is the mask generator model. Typically, CNNs were popular choices when it comes to audio processing, and they still remain in use to date\cite{luo_conv-tasnet_2019, oord_wavenet_2016, ping_deep_2018}. CNNs are excellent at capturing local features, but they are limited by their receptive field. To overcome this, a transformer architecture is added. This allows the model to place attention on relevant parts of the feature space. For instance, to produce a heart sound mask, the transformer model might place its attention on the features that correspond to heart sounds in the feature space. Figure~\ref{fig:mask_generator} shows the implementation of the mask generator block.

The mask generator block starts with layer normalisation and a pointwise 1D convolution to downsample the feature space. This is followed by $N$ 1D convolution blocks to capture local features. These consist of a 1D convolution, followed by a layer normalisation and a Gaussian Error Linear Unit (GELU) activation layer. After that, positional encodings were added before passing it to $s$ transformer encoder blocks that are $M$ layers deep. Next, the outputs from each transformer block are passed through another GELU activation layer and a pointwise 1D convolution producing a mask of size $(F, M)$. Each of the outputs is then stacked together, producing an output mask of size $(s, F, M)$. Finally, the masks are then passed through a ReLU activation to enforce non-negative masks.

\section{Experiments}
\label{sec:experiments}
\subsection{Data Acquisition}
Raw chest sound recordings were obtained from a previous study conducted by Grooby et al \cite{grooby_noisy_2022}, in which further details of the dataset generation can be found. 71 chest sounds were collected from newborn babies admitted to Monash Children’s Hospital with the approval of the Monash Health Human Research Ethics Committee (HREA/18/MonH/471). The one-minute recordings were obtained from the right anterior chest of preterm and term newborns using a CliniCloud digital stethoscope. The recordings were then saved on a smartphone in MP3 format at 16\,kHz sampling rate. The data was downsampled to 4\,kHz sampling rate to reduce data size. 

Separately, 33 chest sound recordings containing synchronous vital signs were also collected and formed the real-world dataset. The synchronous vital signs include second-by-second heart rate from electrocardiogram data and breathing rate from thoracic impedance data.

\subsection{Artificial Dataset}
\begin{figure}[ht]
  \includegraphics[width=\linewidth]{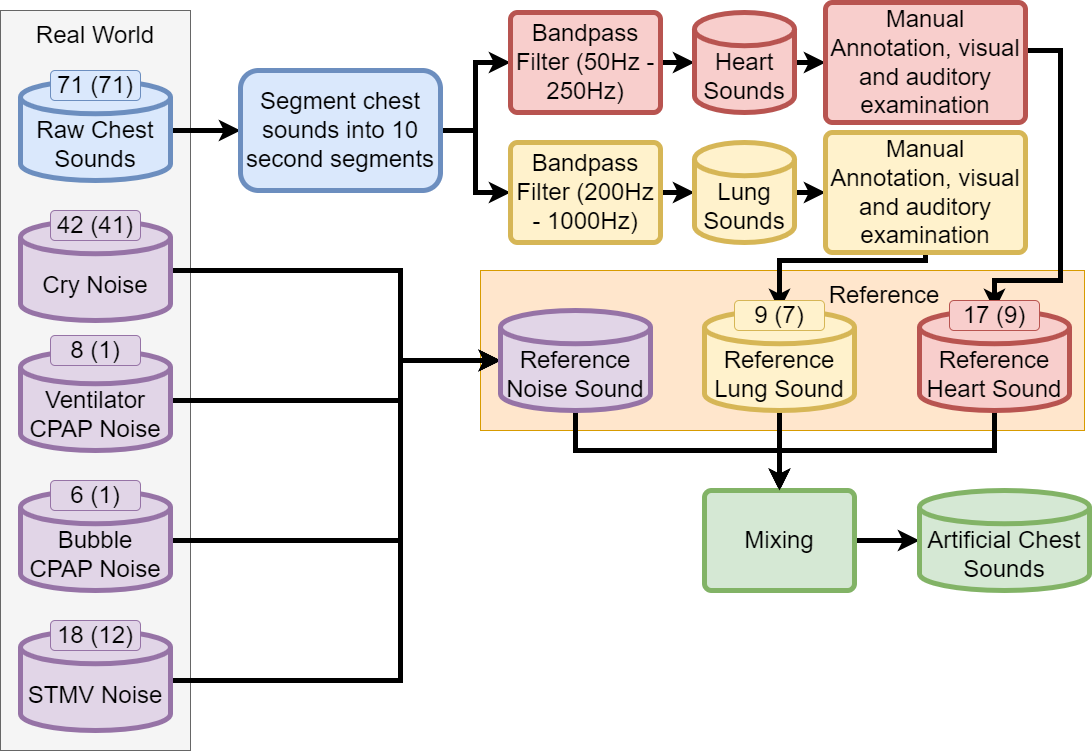}
  \caption{Flowchart for constructing the artificial chest sounds. Each database is labelled with the number of recordings and number of subjects in brackets. STMV Noise here refers to stethoscope movement noises.}
  \label{fig:data_processing}
\end{figure}

Figure~\ref{fig:data_processing} summarises the process of obtaining the artificial chest sounds from the real-world chest sounds.

The raw chest sounds were obtained from recordings in ideal conditions without any respiratory support. The raw chest sounds were first segmented into 10-second recordings. Each segment is then passed through a 4\textsuperscript{th} order Butterworth bandpass filter with passband frequencies of 50-250\,Hz to obtain heart sounds and passed through a 4\textsuperscript{th} order Butterworth bandpass filter with passband frequencies of 200-1000\,Hz to obtain lung sounds. The filtered sounds were then ranked by 3 clinicians and 4 electrical engineers familiar with biomedical auscultation from 1 to 5, with 1 being noisy and hardly detectable and 5 being clear heart/lung sounds. Mean scores of 4 and above were assessed visually and auditorily, and only recordings with strong heart and lung sounds with little to no noise were chosen to form the reference heart and lung sounds.

The reference respiratory support noises such as continuous positive airway pressure machines (CPAP) were obtained by placing a digital stethoscope on the machine or tubing. In addition, chest recordings with high respiratory noise and minimal other sounds were also added. In total, there are 6 ventilator CPAP noises and 8 bubble CPAP noises.

18 reference stethoscope movement noises were obtained from manual annotations from two clinicians and one electrical engineer familiar with biomedical auscultation in the data collected. The annotations included the presence of stethoscope movement, the volume (low, medium, high), and classifying the movement as disconnect or rubbing. In the case of disconnect, no heart and lung sounds should be present while rubbing may still contain heart and lung sounds. Hence, the reference heart and lung sounds were zeroed out at the time instance during mixing.

Reference cry sounds were obtained from regions in the collected data with at least 10 seconds of crying. The regions were extracted from a cry detection algorithm and passed through a 2\textsuperscript{nd} order Butterworth high pass filter with a cutoff frequency of 300\,Hz. Regions that don't contain crying such as inhalation segments were excluded.

Just before mixing, the reference heart and lung sounds were first partitioned into training, validation, and testing partitions according to the baby that it was collected from, with each baby belonging to only one partition to avoid data leakage. Similar partitioning was done for the cry noise and stethoscope movement noise.

The artificial chest sound dataset is constructed by mixing data from the reference heart sound, reference lung sound, and reference noise. Firstly, the signals were normalised to have the same signal power, then rescaled to different relative signal power ranging from -10\,dB to 10\,dB with a step size of 5\,dB. The mixture is then mixed additively as described in Equation~\ref{eqn: additive_mix} and convolutively as described in Equation~\ref{eqn:conv_mix}, where $\Vec{a}$ are random finite-impulse response filters of length 3 and $\| \Vec{a} \|^2 = 1$. $\ast$ is defined as the convolution operation, where $(a \ast b)(t) = \sum_{k=0}^{k=N-1}{a(k)b(t-k)}$

\begin{equation}\label{eqn: additive_mix}
    s_\text{mixture} = s_{\text{heart}} + s_{\text{lung}} + s_{\text{noise}}
\end{equation}

\begin{equation}\label{eqn:conv_mix}
    s_\text{mixture} = s_{\text{heart}}\ast a_{\text{heart}} + s_{\text{lung}}\ast a_{\text{lung}} + s_{\text{noise}}\ast a_{\text{noise}}
\end{equation}

Since each mixture is 10 seconds long sampled at 4,000\,Hz, the final mixture length, $T$, is 40,000 samples.

\subsection{Training Dataset}
The following modifications were made to the training dataset to improve the model's performance:

\begin{enumerate}
    \item The reference noise sound in the training dataset was first rescaled to have a relative signal power of -20\,dB to 0\,dB to ensure that the model was able to learn to identify heart and lung sounds before increasing the relative signal power to be between -10\,dB to 10\,dB during the fine-tuning phase.
    \item Instead of having a discrete relative signal power scaling for the lung and noise sound, the signal power scaling was randomly sampled in the specified range.
    \item Stethoscope movement noise is taken out of the training dataset as it was found to decrease the overall performance of the reconstructed lung sound. Note that the stethoscope movement noise is still present in the test dataset.
\end{enumerate}

In addition to the changes above, the following changes were also made to increase the diversity of the samples:

\begin{enumerate}
    \item For the convolutive mixtures, a random filter length was chosen between 3 and 5.
    \item Instead of training on the full 10-second segments, an 8-second segment is randomly cropped and trained on. This means the model is only trained on 32,000 length samples instead of the full 40,000 length.
\end{enumerate}

For the convolutive mixtures, it was found that the augmented heart and lung sounds were not auditorily different compared to the original samples, but it generated diverse enough samples to prevent the model from overfitting significantly.

\subsection{Real-world Dataset}
In addition to the artificial dataset, 33 chest sounds were recorded with synchronous data consisting of heart rate measurements from electrocardiograms and breathing rate measurements from electrical impedance tomography sensors. These chest sounds were from a separate set of newborns from the heart sounds used to generate the artificial dataset. These chest sounds were further divided into 21 chest sounds without respiratory support sounds and 12 chest sounds with respiratory support sounds.

\subsection{Model Configuration}
Table~\ref{tab:model_param} summarises the model parameter selected for the model used. The following hyperparameters were found by sweeping through different combinations of hyperparameters and choosing the best-performing one based on the performance evaluation on the validation dataset. The following model configuration forms the baseline model when discussing the effects of different model parameters in Section \ref{subsec:ablative}. 

\begin{table}[ht]
\centering
  \caption{Model parameter used for the final model}
  \label{tab:model_param}
  \begin{tabular}{p{0.5\linewidth}  p{0.1\linewidth}}
   \toprule
Encoder/Decoder    & Value    \\ \midrule
Kernel Size        & 512 \\
Feature Size       & 512 \\ \midrule
Mask Generator     & Value    \\ \midrule
Mask Feature Size  & 256 \\
Conv Kernel Size   & 3   \\
Conv Layers        & 6   \\
Num Heads          & 4   \\
Transformer Layers & 4   \\ \bottomrule
\end{tabular}
\end{table}

\subsection{Training Configuration}
When training our model, the following hyperparameters were selected based on the performance of the model on the validation dataset: (a) the model was trained for 40 epochs, (b) The model was trained using the AdamW optimiser with the AMSGrad extension \cite{phuong_convergence_2019} and a weight decay of 0.1, (c) the model was trained with a learning rate scheduler where an initial learning rate of 1e-4 is used, with the learning rate being scaled by 0.5 when the validation accuracy does not improve for 4 epochs, (d) The gradient in the network is clipped if the L2-norm of the gradient is greater than 5, (e) The objective of the training is to maximise the scale-invariant signal-to-distortion ratio (SI-SDR) between the predicted signals $s_{\text{est}}$ and the target signals $s_{\text{target}}$, defined in Equation~\ref{eqn:sisdr}. 

\begin{equation}\label{eqn:sisdr}
  \begin{gathered}
    \alpha = \frac{s_{\text{est}} \cdot s_{\text{target}}}{\|s_{\text{target}}\|^2}\\
    e_{\text{noise}} = \alpha s_{\text{target}} - s_{\text{est}}\\
    \text{SI-SDR} = 10 \log_{10}{\frac{\|\alpha s_{\text{target}}\|^2}{\|e_{\text{noise}}\|^2}}
  \end{gathered}
\end{equation}
 
\subsection{Performance evaluation}
Signal-to-distortion ratio improvement (SDRi) and scale-invariant SDRi (SI-SDRi) were used as objective measures of the performance of the separation method on the artificial data. SI-SDR is defined in Equation~\ref{eqn:sisdr}, while SDR is defined in Equation~\ref{eqn:sdr_eqn}, where the estimated source can be decomposed as shown in Equation~\ref{eqn:signal_decomp}. Table~\ref{tab:signal_decomp_desc} contains the description of the decomposed signal.

\begin{equation}\label{eqn:signal_decomp}
    s_{\text{est}} = s_{\text{target}} + e_{\text{interf}} + e_{\text{noise}} + e_{\text{artif}}
\end{equation}

\begin{table}[ht]
  \caption{Description of the decomposed estimated signal}
  \label{tab:signal_decomp_desc}
    \begin{tabular}{p{0.2\linewidth}  p{0.7\linewidth}}
    \toprule
    Signal              & Description                                                                                \\
    \midrule
    $s_{\text{est}}$    & Estimated source from the separation algorithm                                             \\
    $s_{\text{target}}$ & Target source with some allowed deformation                                                \\
    $e_{\text{interf}}$ & Allowed deformation of sources which accounts for the interference of the unwanted sources \\
    $e_{\text{noise}}$  & Allowed deformation of the perturbating noise                                              \\
    $e_{\text{artif}}$  & Artifacts of the separation algorithm  \\
    \bottomrule
    \end{tabular}
\end{table}

\begin{equation}\label{eqn:sdr_eqn}
    \text{SDR} = 10 \log_{10}\frac{\|s_{\text{target}}\|^2}{\|e_{\text{interf}} + e_{\text{artif}} + e_{\text{noise}}\|^2}
\end{equation}

SDRi and SI-SDRi were then evaluated on the testing partition of the artificial chest sound. The testing partition is further divided into three partitions depending on the type of noise sound present in the partition: (1) No Noise partition, which is when the input mixture does not contain any external noise. (2) General Noise partition, which is when the input mixture contains crying noises and stethoscope movement noises. (3) Respiratory Support Noise, which is when the input mixture contains bubble CPAP noise and ventilator CPAP noise.

For the 33 real-world unseen data that contain vital signs, the heart rate error improvement and breathing rate error improvement were reported as the difference before and after passing through the model when compared to the vital signs. The heart rate estimation was calculated using the modified version of the method by Springer et al. \cite{springer_logistic_2016}. The breathing rate was calculated from a 300-450\,Hz power spectral envelope every second using peak detection. 

In terms of speed comparison, the models were executed on an Intel Core i7-12800H CPU paired with a Nvidia RTX A1000. For the single instance measurement, the input waveform was generated randomly with a length of 40,000 samples at 4\,kHz, making the sample length 10 seconds. The input mixture was also normalised to have values between -1 and 1. For the batched instance, the input waveform was batched and processed with a batch size of 16. The time taken for computation was then scaled down by the batch size. For the GPU instance measurement, the overhead of putting the data into memory is omitted. Every measurement was done 10 times and the mean time taken was reported. 

\section{Results and Discussion}
\label{sec:results}
\subsection{Objective Distortion Measurement on Artificial Dataset}

\begin{figure*}[ht]
  \centering
  \includegraphics[width=0.8\linewidth]{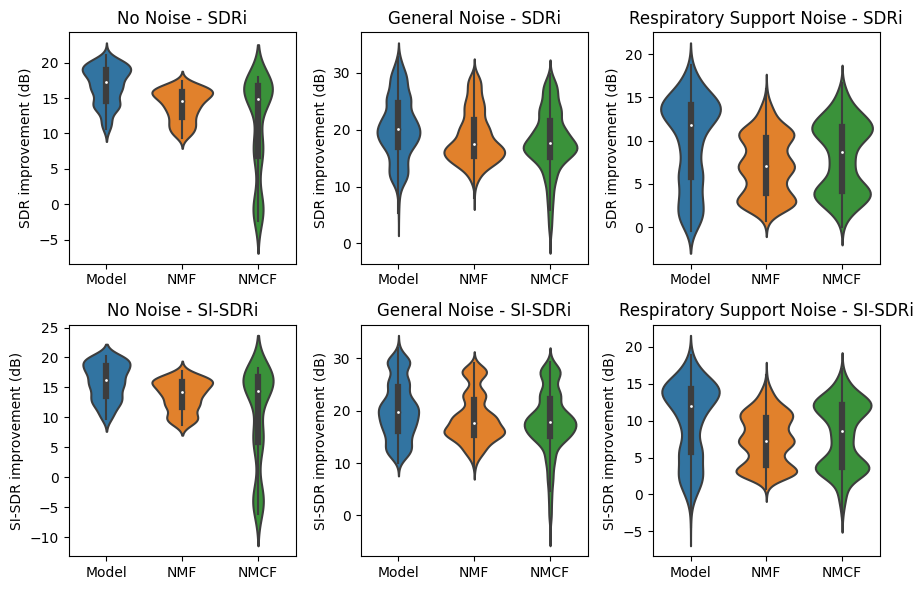}
  \caption{Violin plots of the SDRi and SI-SDRi results for the separated heart sounds. Resp = Respiratory Support.}
  \label{fig:sxr_heart}
\end{figure*}

\begin{figure*}[ht]
  \centering
  \includegraphics[width=0.8\linewidth]{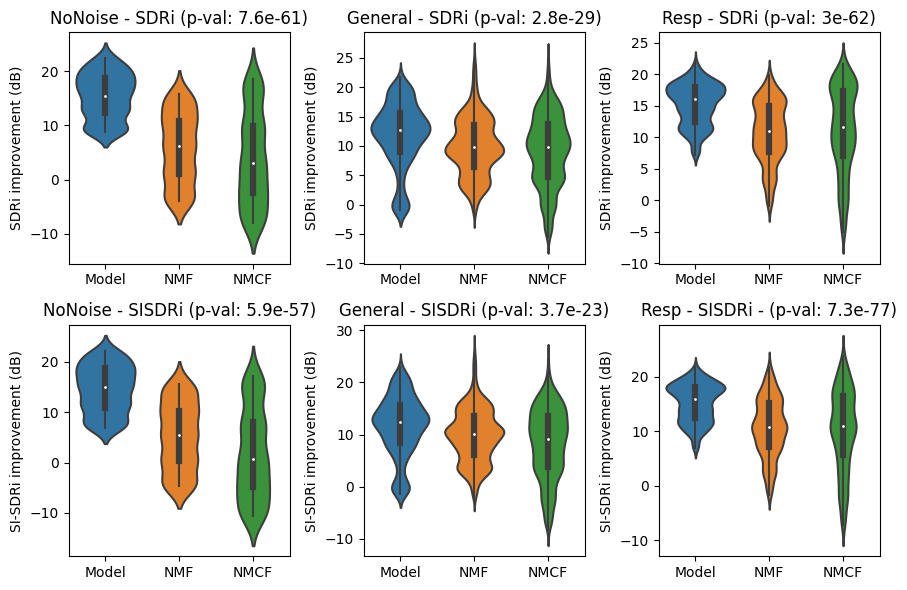}
  \caption{Violin plots of the SDRi and SI-SDRi results for the separated lung sounds. Resp = Respiratory Support.}
  \label{fig:sxr_lung}
\end{figure*}

We investigate the objective measures of the separation method of our model compared to the NMF and NMCF methods that were recently proposed for neonatal chest sound separation. Figure~\ref{fig:sxr_heart} and Figure~\ref{fig:sxr_lung} show the violin plots for the SDRi and SI-SDRi results for each method in separating heart and lung sounds, while Table~\ref{tab:sxr_summary_heart} and Table~\ref{tab:sxr_summary_lung} show the median SDRi and SI-SDRi results for the different methods in separating heart and sounds.

\begin{table*}[ht]
  \caption{Median SDRi and SI-SDRi results for the artificial heart sounds}
  \label{tab:sxr_summary_heart}
  \makebox[\textwidth]{
    \begin{tabular}{lccccccccc}
    \toprule
    Noise Type & \multicolumn{3}{c}{No Noise} &     \multicolumn{3}{c}{General Noise} & \multicolumn{3}{c}{Respiratory Support Noise} \\
    \cmidrule(lr){2-4}\cmidrule(lr){5-7}\cmidrule(lr){8-10}
    Methods & Model & NMF & NMCF & Model & NMF & NMCF &     Model & NMF & NMCF \\
    \midrule
    SDR (dB) & \textbf{17.24} & 14.64 & 14.85 & \textbf{20.16} & 17.42 & 17.57 & \textbf{11.77} & 7.04 & 8.67 \\
    SI-SDR (dB) & \textbf{16.21} & 14.23 & 14.36 & \textbf{19.81} & 17.60 & 17.80 & \textbf{11.99} & 7.30 & 8.53 \\
    \bottomrule
    \end{tabular}
  }
\end{table*}

\begin{table*}[ht]
  \caption{Median SDRi and SI-SDR results for the artificial lung sounds}
  \label{tab:sxr_summary_lung}
  \makebox[\textwidth]{
    \begin{tabular}{lccccccccc}
    \toprule
    Noise Type & \multicolumn{3}{c}{No Noise} &     \multicolumn{3}{c}{General Noise} & \multicolumn{3}{c}{Respiratory Support Noise} \\
    \cmidrule(lr){2-4}\cmidrule(lr){5-7}\cmidrule(lr){8-10}
    Methods & Model & NMF & NMCF & Model & NMF & NMCF &     Model & NMF & NMCF \\
    \midrule
    SDR (dB) & \textbf{15.50} & 6.14 & 3.07 & \textbf{12.71} & 9.81 & 9.72 & \textbf{16.08} & 11.02 & 11.68 \\
    SI-SDR (dB) & \textbf{15.01} & 5.40 & 0.74 & \textbf{12.47} & 10.19 & 9.12 & \textbf{15.90} & 10.79 & 10.93 \\
    \bottomrule
    \end{tabular}
  }
\end{table*}

Our findings demonstrated a consistent enhancement in the objective distortion measurements across all aspects. When it comes to heart sound separation, the proposed model observed modest improvements in the absence of noise and general noise environments, except for the separated lung sounds in the absence of noise, which observed a substantial improvement in the lung sound generated compared to previous methods. However, we observe significant improvements in the presence of respiratory support noise compared to previous methods. This could be down to the advantage of using a convolution-based encoder/decoder architecture rather than using STFT, where the performance of the proposed model regressed to performance similar to the previous NMF and NMCF methods which use STFT. Results of the STFT method can be found in Section \ref{subsec:ablative}.

Whilst modest advancement was observed in the absence of noise and general noise environments, the breakthrough achieved in handling respiratory support noise signifies a major stride forward, addressing previous limitations in our research efforts.

\subsection{Real-world heart-rate and breathing-rate analysis}

We study the effect of applying model separation algorithms to improve the accuracy of the heart-rate estimation algorithm and breathing-rate estimation algorithm. Table~\ref{tab:hr} shows the heart rate improvement and Table~\ref{tab:br} shows the breathing rate improvement for the real-world chest sounds for each separation method when compared to the vital signs collected.

\begin{table}[ht]
  \caption{The mean, median, and standard deviation of heart rate improvement. Nil signified that no respiratory support machines were present during the collection of the chest sounds. Resp signified that respiratory support machines were present during the collection of the chest sounds.}
  \label{tab:hr}
  \centering
  \begin{tabular}{lcccccc}
  \toprule
  Method & \multicolumn{3}{c}{HR improvement - Nil} & \multicolumn{3}{c}{HR improvement - Resp} \\
   \cmidrule(lr){2-4}\cmidrule(lr){5-7}
   & mean & median & std & mean & median & std \\
  \midrule
  Model & \textbf{2.62} & 0.00 & 12.23 & 3.29 & 0.36 & 19.40 \\
  NMF & 1.92 & 0.00 & 12.67 & 2.65 & 0.11 & 11.96 \\
  NMCF & 2.24 & 0.00 & 12.41 & \textbf{5.47} & 0.53 & 18.61 \\
  \bottomrule
  \end{tabular}
\end{table}

\begin{table}[ht]
  \caption{The mean, median, and standard deviation of breathing rate improvement. Nil signified that no respiratory support machines were present during the collection of the chest sounds. Resp signified that respiratory support machines were present during the collection of the chest sounds.}
  \label{tab:br}
  \centering
  \begin{tabular}{lcccccc}
  \toprule
  Method & \multicolumn{3}{c}{BR improvement - Nil} & \multicolumn{3}{c}{BR improvement - Resp} \\
   \cmidrule(lr){2-4}\cmidrule(lr){5-7}
   & mean & median & std & mean & median & std \\
  \midrule
  Model & \textbf{1.89} & 0.00 & 6.45 & \textbf{0.10} & 0.00 & 6.15 \\
  NMF & 1.45 & 0.00 & 10.44 & -1.04 & 0.00 & 8.01 \\
  NMCF & 1.27 & 0.00 & 11.64 & -1.30 & 0.00 & 8.43 \\
  \bottomrule
  \end{tabular}
\end{table}

When observing the heart rate improvement, the results were mixed. When it comes to the no respiratory support case, the proposed model performed better, and when it comes to the respiratory support case, the previous NMCF method seemed to perform better. However, the proposed model seemed to perform better in terms of breathing rate improvement compared to the previous two methods in both scenarios. It should be noted that the median of all the metrics was close to 0, indicating that most of the results are positively skewed. This result could be down to a few factors: (a) In the heart rate improvement case, the heart rate estimation algorithm is already robust to noises, and all three algorithms only work to improve the outlier samples. This observation is also supported when looking at the predicted heart rate before separation already being close to the true heart rate. (b) The chest sound quality for most of the samples was low, with minimal to nonexistent detection of both heart and lung sounds. This is especially true for the respiratory support samples, where the chest sound recordings were dominated by the respiratory support machine noises. Notwithstanding the foregoing, these samples are typically what is expected, and further improvement has to be made here to improve the performance of these models.

\subsection{Computation Time}

We analyse the computation times for different methods. Table~\ref{tab:computation_time} shows the computational time of the proposed method compared to the previous methods.

\begin{table}[ht]
  \caption{Computation time comparisons. All measurements were made in milliseconds except for the NMCF method, which is made in seconds.}
  \label{tab:computation_time}
  \centering
  \begin{tabular}{@{}lccc@{}}
    \toprule
    Method                & Model    & NMF      & NMCF    \\ \midrule
    Single Instance       & 42.04\,ms & 714.0\,ms & 23.92\,s \\
    Batch Instance        & 18.42\,ms & N/A      & N/A     \\ \midrule
    Single Instance (GPU) & 23.88\,ms & N/A      & N/A     \\
    Batch Instance (GPU)  & 1.320\,ms & N/A      & N/A     \\ \bottomrule
  \end{tabular}
\end{table}

It can be seen that the proposed model is significantly faster than the previous methods. For the single instance case, the proposed model is 17 times faster than the NMF method and 570 times faster than the NMCF method. Not only that, the proposed model has batch processing and GPU support, further increasing the speed compared to the previous methods. However, it is important to note that the previous NMF and NMCF methods were implemented in MATLAB, and more optimisation could be done to improve the computation time. Thus, the proposed model could be useful for real-time applications.

\subsection{Model and Training Optimisation}\label{subsec:ablative}
We study the effects of different model parameters and training configurations on the performance of the model. Table~\ref{tab:ablative} shows the objective distortion measurements for different modifications made. The baseline model is the same model described in Table~\ref{tab:model_param}. From the table, we can conclude the following:

\begin{table*}[htbp]
  \caption{Effect of different model configurations. The changes were made from the baseline model configuration in Table \ref{tab:model_param}.}
  \label{tab:ablative}
  \centering
  \begin{tabular}{p{0.4\textwidth}ccccc}
    \toprule
    Changes & Properties & \multicolumn{4}{c}{Metric} \\
    \cmidrule(lr){2-2}\cmidrule(lr){3-6}
    & Model Size & SDR Heart & SDR Lung & SI-SDR Heart & SI-SDR Lung \\
    & (M) & (dB) & (dB) & (dB) & (dB) \\
    \midrule
    Baseline & 8.42 & \textbf{16.39} & 14.76 & \textbf{16.00} & 14.46 \\
    \midrule
    Removed conv in mask generator, added transformer layers to compensate for smaller model size & 10.40 & 14.25 & 14.32 & 13.91 & 13.96 \\
    Changed encoder/decoder to STFT & 7.90 & 13.31 & 14.39 & 13.01 & 14.20 \\
    Decrease encoder kernel size to 256 & 8.16 & 15.61 & \textbf{15.04} & 15.39 & \textbf{14.57} \\
    Increase encoder kernel size to 1024 & 8.95 & 16.05 & 12.55 & 15.51 & 11.63 \\
    Decrease feature space size to 256 & 7.96 & 15.95 & 14.75 & 15.52 & 14.08 \\
    Increase feature space size to 1024 & 9.34 & 15.80 & 14.37 & 15.47 & 14.05 \\
    Trained without random cropping & 8.42 & 11.57 & 6.35 & 11.24 & 5.87 \\
    Trained with relative SNR noise from -10\,dB to 10\,dB & 8.42 & 15.97 & 13.35 & 15.67 & 12.57 \\
    Trained with stethoscope movement & 8.42 & 16.32 & 8.33 & 16.00 & 7.44 \\
    \bottomrule
  \end{tabular}
\end{table*}

\begin{enumerate}
  \item The convolution model before the transformer is important to improve the performance of the model.
  \item The use of convolution/transposed convolution for the encoder/decoder pair improves the performance of the model, especially in the respiratory support noise cases, where the performance of the model (SDR-Heart: 5.66\,dB, SDR-Lung: 13.21\,dB, SI-SDR-Heart: 5.86\,dB, SI-SDR-Lung: 13.46\,dB) is similar to the previous methods which also uses STFT.
  \item A smaller kernel size has a small improvement to the performance of the lung sound created. This, however, comes at the cost of the quality of the heart sounds created.
  \item An optimal model performance is achieved with a feature size of 512.
  \item Not using random cropping during training led to the model overfitting, as indicated by a notable 5\,dB difference between training and validation loss, along with a gradual rise in validation loss towards the end of training. Consequently, the model's performance is subpar. It's worth noting that the model selected is based on validation loss, not the final model.
  \item Training with higher relative SNR noise from the start to decrease the performance of the model. This could be down to the model not being able to learn in a high-noise environment.
  \item Training with the stethoscope movement noise seemed to decrease the performance of the lung sound created. 
\end{enumerate}

\subsection{Output Waveform Analysis}

\begin{figure}[ht]
  \includegraphics[width=\linewidth]{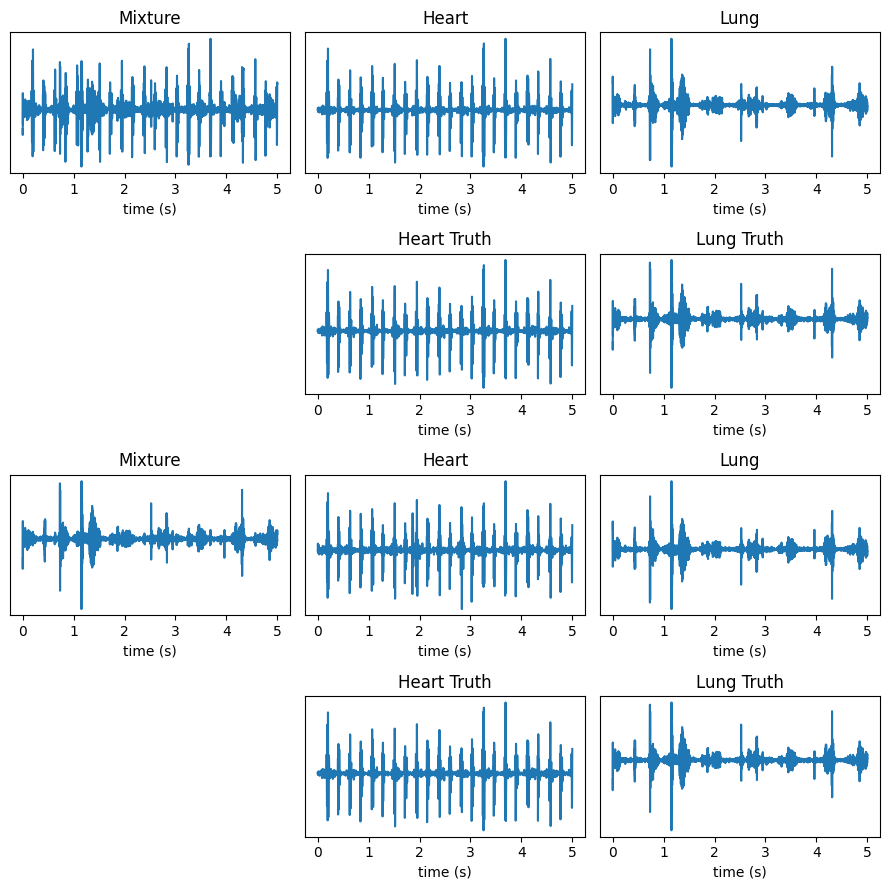}
  \caption{Example waveforms generated. Mixture = mixture of heart and lung sounds. Heart = heart sound separated by the model. Lung = lung sound separated by the model. Heart Truth and Lung Truth = heart and lung sounds used to generate the mixture.}
  \label{fig:waveform}
\end{figure}

We investigate the waveforms generated by the model. Figure~\ref{fig:waveform} shows some waveform generated by the proposed model when the mixture contains different ratios of heart and lung sounds. 

Overall, the model performs well at separating mixed heart and lung sounds, regardless of their power. While the separated lung sound still contains some spikes from the heart sounds, this is mainly an artifact of the ground truth lung sounds also containing the same spikes.

\begin{figure}[ht]
  \includegraphics[width=\linewidth]{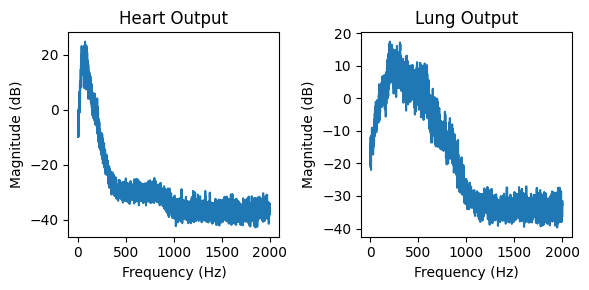}
  \caption{Fourier transform out the output heart sound and output lung sound.}
  \label{fig:output_fft}
\end{figure}

Figure~\ref{fig:output_fft} shows the frequency content of the output waveform, where the output is the Fourier transform of the output heart and lung sound. From the figure, it is observed that the output heart sound has a narrow frequency content centred around 150\,Hz. This matches our understanding of heart sounds. The lung output, on the other hand, has a wider frequency range where we expect the breathing information to reside. Therefore, the recreated heart and lung sounds fall within the expected range of operation.

\subsection{Future works}

Although the model performed well with artificial chest sound mixtures, there is still room for improvement in its real-world chest sound performance. A basic idea here is to incorporate real-world metrics such as heart error rate, breathing error rate, or subjective signal-quality metrics into the loss function to improve the performance of the model for real-world use.

Another improvement that could be made is in the ground truth labels. Currently, the ground truth lung sounds still contain some heartbeats due to our methods of obtaining the heart and lung sounds. An improvement that could be made here is to use physics-informed neural networks to extract heart and lung sounds. That way, heart and lung sounds that are not governed by a physical model will be harshly penalised, potentially removing some of the spikes generated in the lung sounds.

\section{Conclusion}
\label{sec:conclusion}
We conclude that the proposed deep learning-based sound separation method represents an advancement in neonatal chest sound separation compared to previous methods. These improvements suggest that the proposed model could potentially replace previous neonatal chest sound separation methods. For example, the improved objective distortion measurements from our model imply that the separated heart and lung sounds are of better quality than previous attempts, potentially making them suitable as a preprocessing step for various algorithms involving phonocardiogram-based health monitoring systems. Additionally, the significantly lower computational costs suggest that the proposed model could be suitable for real-time applications. Furthermore, subjective signal-quality measurements and the exploration of a physics-informed neural network remain uncharted territory, which may help bridge the gap between real-world chest sound separation and the removal of noisy ground truth samples.

\section*{Acknowledgment}
We thank and acknowledge the support from Monash Newborn and Monash Children's Hospital for the data provided.

\printbibliography
\end{document}